\documentclass[pra,showpacs,twocolumn]{revtex4-1}
\usepackage{amsfonts}
\usepackage{amsmath}
\usepackage{amssymb,amscd}
\usepackage{psfrag}
\usepackage{graphicx}
\usepackage{braket}
\usepackage{epsfig}
\usepackage{epsfig}
\usepackage{subfigure}
\usepackage{color}
\usepackage{amssymb}

\begin{document}

\title{Phase estimation of phase shifts in two arms for an SU(1,1)
interferometer with coherent and squeezed vacuum states}
\author{Qian-Kun Gong$^{1}$}
\author{Dong Li$^{1}$}
\author{Chun-Hua Yuan$^{1,4}$}
\email{chyuan@phy.ecnu.edu.cn}
\author{Z. Y. Ou$^{1,3}$}
\author{Weiping Zhang$^{2,4}$}

\address{$^1$Quantum Institute for Light and Atoms, Department of Physics, East China
Normal University, Shanghai 200062, P. R. China}

\address{$^2$Department of Physics and Astronomy, Shanghai Jiao Tong
University, Shanghai 200240, P. R. China}

\address{$^3$Department of Physics, Indiana University-Purdue University
Indianapolis, 402 North Blackford Street, Indianapolis, Indiana 46202, USA}

\address{$^4$Collaborative Innovation Center of Extreme Optics, Shanxi
University, Taiyuan, Shanxi 030006, P. R. China}

\begin{abstract}
We theoretically study the quantum Fisher information (QFI) of the SU(1,1)
interferometer with phase shifts in two arms by coherent $\otimes$ squeezed
vacuum state input, and give the comparison with the result of phase shift
only in one arm. Different from the traditional Mach-Zehnder interferometer,
the QFI of single-arm case for an SU(1,1) interferometer can be slightly
higher or lower than that of two-arm case, which depends on the intensities
of the two arms of the interferometer. For coherent $\otimes$ squeezed
vacuum state input with a fixed mean photon number, the optimal sensitivity
is achieved with a squeezed vacuum input in one mode and the vacuum input in
the other.
\end{abstract}

\maketitle

\section{Introduction}

Quantum enhanced metrology which has received a lot of attention in recent
years is the use of quantum measurement techniques to obtain higher
statistical precision than purely classical approaches \cite%
{Helstrom76,Holevo82,Caves81,Braunstein94,Braunstein96,Lee02,Giovannetti06,Zwierz10,Giovannetti04,Giovannetti11,Ou,Abbott,Hosten16,Xiang,Zhang,Wei}%
. Mach-Zehnder interferometer (MZI) and its variants were used as a generic
model to realize high precise estimation of phase. In order to achieve the
ultimate lower bounds \cite{Zwierz,Luca2015}, much work has been devoted to
find the methods to improve the sensitivity of phase estimation, such as (1)
using the nonclassical input states (quantum resources)-squeezed states \cite%
{Caves81,Xiao,Grangier} and NOON states\cite{Dowling08,NOON}; (2) using the
new detection methods-homodyne detection\cite{Li14,Hu} and parity detection
\cite{Anisimov,Gerry2010,Chiruvelli,Li2016}; (3) using the nonlinear
processes-amplitude amplification \cite{Yurke86} and phase magnification
\cite{Hosten16}. Here we focus on the nonlinear amplitude amplification
process to improve the sensitivity. In 1986, Yurke \textit{et al.} \cite%
{Yurke86} introduced a new type of interferometer where two nonlinear beam
splitters (NBSs) take the place of two linear beam splitters (BSs) in the
traditional MZI. It is also called the SU(1,1) interferometer because it is
described by the SU(1,1) group, as opposed to the traditional SU(2) MZI for
BS. The detailed quantum statistics of the two-mode SU(1,1) interferometer
was studied by Leonhardt \cite{Leonhardt}. The SU(1,1) phase states were
also studied theoretically in quantum measurements for phase-shift
estimation \cite{Vourdas,Sanders}. Furthermore, the SU(1,1)-type
interferometers have been reported by different groups using different
systems in theory and experiment, such as all optical arms\cite%
{Hudelist,Lett,Plick,Marino}, all atomic arms\cite%
{Linnemann16,Gabbrielli,Gross}, atom-light hybrid arms\cite%
{ChenPRL15,Chen16,Yama,Haine,Szigeti,Haine16}, light-circuit quantum
electrodynamics system hybrid arms \cite{Barzanjeh}, and all mechanical
modes arms\cite{Cheung}. These SU(1,1)-type interferometers provide
different methods for basic measurement.

At present, many researchers are focusing on how to measure the phase
sensitivities, where several detection schemes have been presented\cite%
{Li14,Li2016,Marino}. In general, it is difficult to optimize all the
detection schemes to obtain the optimal estimation protocol. However, the
quantum Fisher information (QFI) \cite{Braunstein94,Braunstein96}
characterizes the maximum amount of information that can be extracted from
quantum experiments about an unknown parameter (e.g., phase shift $\phi $)
using the best (and ideal) measurement device. Therefore, the lower bounds
in quantum metrology can be obtained by using the method of the QFI. In
recent years, many efforts were made to obtain the QFI of different measure
systems \cite{Toth, PezzBook, Demkowicz,Wang,Jarzyna, Monras, Pinel, Liu13,
Gao,Jiang,Yan15,Safranek,Sparaciari15,Ren16,Sparaciari,Strobel,Lu,Hauke,Liu17}%
. For the SU(1,1) interferometers with phase shift only in one arm, the QFI
with coherent states input was studied by Sparaciari \emph{et al.} \cite%
{Sparaciari15,Sparaciari}, and the QFI with coherent $\otimes $ squeezed
vacuum state input was presented by some of us \cite{Li2016}. Nevertheless
in some measure schemes, the phase shifts in two arms are required to
measure. For example, the phase sensitivity of phase shifts in two arms for
the SU(1,1) interferometer with coherent states input was experimentally
studied by Linnemann \emph{et al.} \cite{Linnemann16}. Jarzyna \emph{et al.}
studied the QFIs of phase shifts in the two-arm case for a MZI, and
presented the relation with the result of phase shift in the single-arm case
\cite{Jarzyna}. Since phase shift in the single arm is not simply equivalent
to that phase shifts in two arms where one phase shift of them is $0$, the
QFIs of phase shifts in two arms for an SU(1,1) interferometer are needed to
research. In this paper, we study the QFI of SU(1,1) interferometer of phase
shifts in two arms with two coherent states input and coherent $\otimes $
squeezed vacuum state input, and give the comparison with the result of
phase shift only in one arm. These results should provide useful help to
some phase measurement processes.

The remaining part of this paper is organized in the following way. In
Section 2 we firstly give a brief review of the SU(1,1) interferometer, then
derive the QFI of phase shifts in two arms for an SU(1,1) interferometer. In
Section 3 the phase sensitivities of SU(1,1) interferometer obtained from
the quantum Cram\'{e}r-Rao bound (QCRB) \cite{Helstrom76,Holevo82} are
discussed, and the results of phase shifts in different arms are compared.
The conclusions are summarized in Section 4.

\begin{figure}[ptbh]
\centerline{\includegraphics[scale=0.5,angle=0]{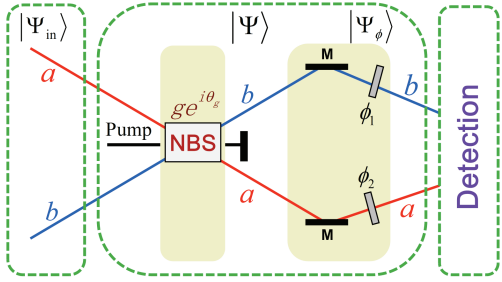}}
\caption{ Schematic diagram of the parameter estimation process based on the
SU(1,1) interferometer. $g$ and $\protect\theta_{g}$ describe the strength
and phase in the first NBS process, respectively. $a$ and $b$ denote two
light modes in the interferometer. In the Schr\"{o}dinger picture the
initial state $|\Psi _{\mathrm{in}}\rangle $ injecting into a NBS results in
the output $|\Psi \rangle$, and the state is transformed as $|\Psi_{\protect%
\phi} \rangle$ after phase shifts. NBS: nonlinear beam splitter; $\protect%
\phi_{1}$, $\protect\phi_{2}$: phase shifts; $\mathrm{M}$: mirrors.}
\label{fig1}
\end{figure}

\section{The QFI of phase shifts in two arms for an SU(1,1) interferometer}

Because the QFI $\mathcal{F}$ is the intrinsic information in the quantum
state and is not related to actual measurement procedure as shown in Fig.~1.
It establishes the best precision that can be attained with a given quantum
probe \cite{Braunstein94,Braunstein96}. In this section, we study the QFIs
of SU(1,1) interferometer of phase shifts in two arms, and compare them with
the results of phase shift only in one arm.

\subsection{NBS and phase shifts}

In an SU(1,1) interferometer, the NBSs take the place of the BSs in the
traditional MZI shown in Fig.~1. Firstly, we theoretically describe the NBS
briefly, which can be completed by the optical parameter amplifier (OPA) or
four-wave mixing (FWM) process. The annihilation operators of the two modes $%
a$, $b$ and the pump field are $\hat{a}$, $\hat{b}$, and $\hat{c}$,
respectively. The interaction Hamiltonian for NBS is of the form
\begin{equation}
\hat{H}=i\hbar \eta ^{\ast }\hat{a}\hat{c}^{\dagger }\hat{b}\hat{c}^{\dagger
}-i\hbar \eta \hat{a}^{\dagger }\hat{c}\hat{b}^{\dagger }\hat{c}.
\end{equation}%
Because the pump field is very strong and the intensity of the pump field is
not significantly changed in the mixing process. Then the initial and final
states of the pump field are the same as the coherent state $|\alpha _{%
\mathrm{pump}}\rangle $. Under the undepleted pump approximation, the
Hamiltonian is written as%
\begin{equation}
\hat{H}=i\hbar \eta ^{\ast }\alpha _{\mathrm{pump}}^{\ast 2}\hat{a}\hat{b}%
-i\hbar \eta \alpha _{\mathrm{pump}}^{2}\hat{a}^{\dagger }\hat{b}^{\dagger }.
\end{equation}%
The corresponding time-evolution operator is $\hat{U}(t)=e^{-i\hat{H}t/\hbar
}=\exp (-\xi \hat{a}^{\dagger }\hat{b}^{\dagger }+\xi ^{\ast }\hat{a}\hat{b}%
) $, where $\xi =\eta \alpha _{\mathrm{pump}}^{2}t=ge^{i\theta _{g}}$ is the
two-mode squeezing parameter. In the Schr\"{o}dinger picture the initial
state $|\Psi _{\mathrm{in}}\rangle $ injecting into a NBS results in the
output $|\Psi \rangle =\hat{U}(t)|\Psi _{\mathrm{in}}\rangle $, where the
transformation of the annihilation operators is%
\begin{equation}
\left(
\begin{array}{c}
\hat{a}^{\prime } \\
\hat{b}^{^{\dagger }\prime }%
\end{array}%
\right) =\left(
\begin{array}{cc}
\cosh g & -e^{i\theta _{g}}\sinh g \\
-e^{-i\theta _{g}}\sinh g & \cosh g%
\end{array}%
\right) \left(
\begin{array}{c}
\hat{a} \\
\hat{b}^{^{\dagger }}%
\end{array}%
\right) .  \label{transf}
\end{equation}

Secondly, we describe the phase shifts process. Different from the BS, the
NBS involves three light fields where the pump field is classical and with a
classical reference phase. The uncertainty of classical pump field $1/\sqrt{%
|\alpha _{\mathrm{pump}}|^{2}}$ is very small and the phase uncertainties
are from the modes $a$ and $b$. After the first NBS, as shown in Fig.~1, the
two beams sustain phase shifts, i.e., the mode $a$ and mode $b$ undergo the
phase shifts of $\phi _{1}$ and $\phi _{2}$, respectively. Then we may write%
\begin{eqnarray}
&&\exp \left( i\phi _{1}\hat{a}^{\dagger }\hat{a}\right) \exp \left( i\phi
_{2}\hat{b}^{\dagger }\hat{b}\right)   \notag \\
&=&\exp \left[ i\frac{\phi _{1}+\phi _{2}}{2}(\hat{a}^{\dagger }\hat{a}+\hat{%
b}^{\dagger }\hat{b})\right] \exp \left[ i\frac{\phi _{1}-\phi _{2}}{2}(\hat{%
a}^{\dagger }\hat{a}-\hat{b}^{\dagger }\hat{b})\right]   \notag \\
&=&\exp \left[ i\frac{\phi }{\hbar }\hat{K}_{z}-i\frac{\phi }{2}\right] \exp %
\left[ i\frac{(\phi _{1}-\phi _{2})}{\hbar }\hat{J}_{z}\right] ,
\end{eqnarray}%
where $\phi =\phi _{1}+\phi _{2}$, $\hat{K}_{z}=\hbar (\hat{a}^{\dagger }%
\hat{a}+\hat{b}^{\dagger }\hat{b}+1)/2$, and $\hat{J}_{z}=\hbar (\hat{a}%
^{\dagger }\hat{a}-\hat{b}^{\dagger }\hat{b})/2$. In the Schr\"{o}dinger
picture the transformation of the incoming state vector $\left\vert \Psi
\right\rangle $ is given as following
\begin{equation}
|\Psi _{\phi }\rangle =e^{-i\phi /2}e^{i(\phi _{1}-\phi _{2})/\hbar \hat{J}%
_{z}}e^{-i\phi /\hbar \hat{K}_{z}}\left\vert \Psi \right\rangle .
\label{phase shift}
\end{equation}%
$\hat{J}_{z}$ is an invariant for the four-wave mixing process. The operator
$e^{i(\phi _{1}-\phi _{2})/\hbar \hat{J}_{z}}$ gives rise to phase factors
which does not contribute to the expectation values of number operators.

\begin{figure}[ptbh]
\centerline{\includegraphics[scale=0.4,angle=0]{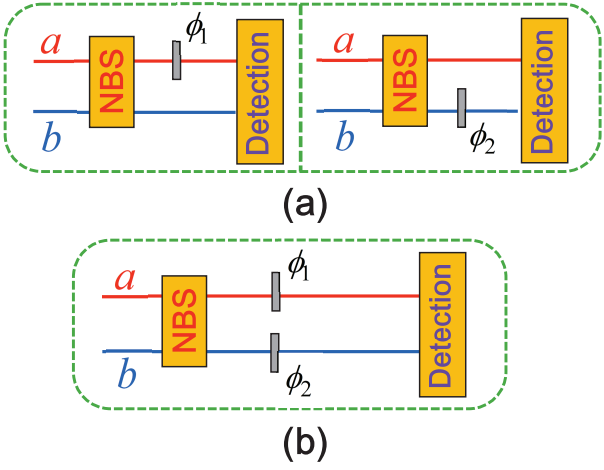}}
\caption{ Different phase delay ways of the interferometer: (a) Phase shift
in the single arm is divided into single upper arm and single lower arm due
to the intensities of the two arms are not equal. (b) Phase shifts in the
two arms.}
\label{fig2}
\end{figure}

\subsection{QFI}

The QFI is the intrinsic information in the quantum state and it is not
related to actual measurement procedure, and is at least as great as the
classical Fisher information for the optimal observable. The QFI $\mathcal{F}
$ is defined as \cite{Braunstein94,Braunstein96}%
\begin{equation}
\mathcal{F}=\mathrm{Tr}[\rho (\phi )L_{\phi }^{2}],
\end{equation}%
where the Hermitian operator $L_{\phi }$, called symmetric logarithmic
derivative, is defined as the solution of the equation $\partial _{\phi
}\rho (\phi )=[\rho (\phi )L_{\phi }+L_{\phi }\rho (\phi )]/2$. In terms of
the complete basis $\{|k\rangle \}$ such that $\rho (\phi
)=\sum_{k}p_{k}|k\rangle \langle k|$ with $p_{k}\geq 0$ and $\sum_{k}p_{k}=1$%
, the QFI can be written as \cite{Braunstein94,Braunstein96,Toth, PezzBook,
Demkowicz}
\begin{equation}
\mathcal{F}=\sum_{k,k^{\prime }}\frac{2}{p_{k}+p_{k^{\prime }}}\left\vert
\langle k|\partial _{\phi }\rho (\phi )|k^{\prime }\rangle \right\vert ^{2}.
\end{equation}%
Under the condition of lossless, for a pure state the QFI is reduced to \cite%
{Jarzyna,PezzBook}%
\begin{equation}
\mathcal{F}=4\left( \langle \Psi _{\phi }^{\prime }|\Psi _{\phi }^{\prime
}\rangle -\left\vert \langle \Psi _{\phi }^{\prime }|\Psi _{\phi }\rangle
\right\vert ^{2}\right) ,  \label{QFI}
\end{equation}%
where $|\Psi _{\phi }^{\prime }\rangle =\partial |\Psi _{\phi }\rangle
/\partial \phi $. In general, the QFI bounds depend on the ways that the
interferometer phase delay is modeled: (I) phase shift only in the single
arm, and (II) phase shifts distributed in two arms, which is shown in
Fig.~2(a) and (b), respectively. Hereafter, we use the single-arm case (S)
and two-arm case (T) to denote them.

Now, we give the QFIs with different input states under the condition of
phase shifts in two arms. From Eq.~(\ref{phase shift}), $|\Psi _{\phi
}\rangle $ is the state vector just before the detection process\ of the
SU(1,1) interferometer and $|\Psi _{\phi }^{\prime }\rangle =-i(1/2+\hat{K}%
_{z}/\hbar )\left\vert \Psi _{\phi }\right\rangle $. Then from Eq.~(\ref{QFI}%
) the QFI can be worked out:%
\begin{equation}
\mathcal{F}=\frac{4}{\hbar ^{2}}\Delta ^{2}\hat{K}_{z},
\end{equation}%
where $\Delta ^{2}\hat{K}_{z}=\langle \Psi |\hat{K}_{z}^{2}|\Psi \rangle
-\langle \Psi |\hat{K}_{z}|\Psi \rangle ^{2}$. Using the transforms of Eqs. (%
\ref{transf}) and (\ref{phase shift}) and with two coherent states $|\alpha
\rangle \otimes |\beta \rangle $ ($j=\left\vert j\right\vert e^{i\theta _{j}}
$, $N_{j}=\left\vert j\right\vert ^{2}$, $j=\alpha $, $\beta $) input case,
for the SU(1,1) interferometer we have
\begin{eqnarray}
\mathcal{F}_{\mathrm{coh\&coh}}^{\mathrm{T}} &=&(N_{\alpha }+N_{\beta
})\cosh (4g)+\sinh ^{2}(2g)  \notag \\
&&+2\sinh (4g)\sqrt{N_{\alpha }N_{\beta }}\cos (\theta _{\alpha }+\theta
_{\beta }-\theta _{g}-\pi ).\notag \\
\end{eqnarray}%
When $\theta _{\alpha }+\theta _{\beta }-\theta _{g}=\pi $, the maximal QFI $%
\mathcal{F}_{\mathrm{coh\&coh}}^{\mathrm{T}}$ is reduced to
\begin{equation}
\mathcal{F}_{\mathrm{coh\&coh}}^{\mathrm{T}}=(N_{\alpha }+N_{\beta })\cosh
(4g)+\sinh ^{2}(2g)+2\sinh (4g)\sqrt{N_{\alpha }N_{\beta }}.  \label{cohcoh}
\end{equation}%
When $N_{\alpha }=N_{\beta }=0$ (vacuum input) and $N_{\alpha }\neq 0$, $%
N_{\beta }=0$ (one coherent state input), from Eq.~(\ref{cohcoh}) the
corresponding QFIs are given by $\mathcal{F}_{\mathrm{vac}}^{\mathrm{T}%
}=\sinh ^{2}(2g)$ and $\mathcal{F}_{\mathrm{coh\&vac}}^{\mathrm{T}%
}=N_{\alpha }\cosh 4g+\sinh ^{2}(2g)$, respectively.

Next, we consider a coherent light combined with a squeezed vacuum light as
the input $|\psi _{in}\rangle =|\alpha \rangle _{a}\otimes |0,\varsigma
\rangle _{b}$ ($\alpha =\left\vert \alpha \right\vert e^{i\theta _{\alpha }}$%
, $N_{\alpha }=\left\vert \alpha \right\vert ^{2}$, and $|0,\varsigma
\rangle _{b}=\hat{S}_{b}(r)|0\rangle _{b}$ is the single-mode squeezed
vacuum state in the $b$-mode where $\hat{S}_{b}(r)=\exp [(\varsigma ^{\ast }%
\hat{b}^{2}-\varsigma \hat{b}^{\dag 2})/2]$ with $\varsigma =r\exp (i\theta
_{\varsigma })$ is the single-mode squeezing parameter), and the QFI can be
worked out:%
\begin{eqnarray}
\mathcal{F}_{\mathrm{coh\&squ}}^{\mathrm{T}} &=&\cosh ^{2}(2g)\left[ \frac{1%
}{2}\sinh ^{2}(2r)+N_{\alpha }\right]  \\
&&+\sinh ^{2}(2g)[N_{\alpha }(\cosh 2r-\sinh 2r\cos \Phi )+\cosh ^{2}r],
\end{eqnarray}%
where $\Phi =\theta _{\varsigma }+2\theta _{\alpha }-2\theta _{g}$. When $%
\Phi =\pi $, the maximal QFI $\mathcal{F}_{\mathrm{coh\&squ}}^{\mathrm{T}}$
is given by
\begin{eqnarray}
\mathcal{F}_{\mathrm{coh\&squ}}^{\mathrm{T}} &=&\cosh ^{2}(2g)\left[ \frac{1%
}{2}\sinh ^{2}(2r)+N_{\alpha }\right]   \notag \\
&&+\sinh ^{2}(2g)[N_{\alpha }e^{2r}+\cosh ^{2}r].
\end{eqnarray}%
When $r=0$, $\mathcal{F}_{\mathrm{coh\&squ}}^{\mathrm{T}}$ is also reduced
to $\mathcal{F}_{\mathrm{coh\&vac}}^{\mathrm{T}}$, which agrees with the
above result. This input state was also used to improve the phase-shift
measurement sensitivity in the SU(1,1) interferometer but only with the
method of the error propagation in Ref.~ \cite{Li14}.

So far, we have given the QFI of SU(1,1) interferometer where the phase
shifts in the two arms, and they as well as the QFIs with phase shift in the
one arm case are summarized in the Table I. From this Table, the QFIs of
phase shift in upper arm and in lower arm are also slightly different
because the intensities in two arms of the interferometer are unbalanced.
The QFI of single-arm case for an SU(1,1) interferometer can be slightly
higher or lower than that of double arms case, which depends on the
intensities of the two arms of the interferometer. Different from the
SU(1,1) interferometer, the QFIs of the phase shifts in single upper arm and
in single lower arm are the same due to the intensity balance of the two
arms for the MZI \cite{Jarzyna}.

\section{QCRB}

\begin{figure}[ptbh]
\centerline{\includegraphics[scale=0.5,angle=0]{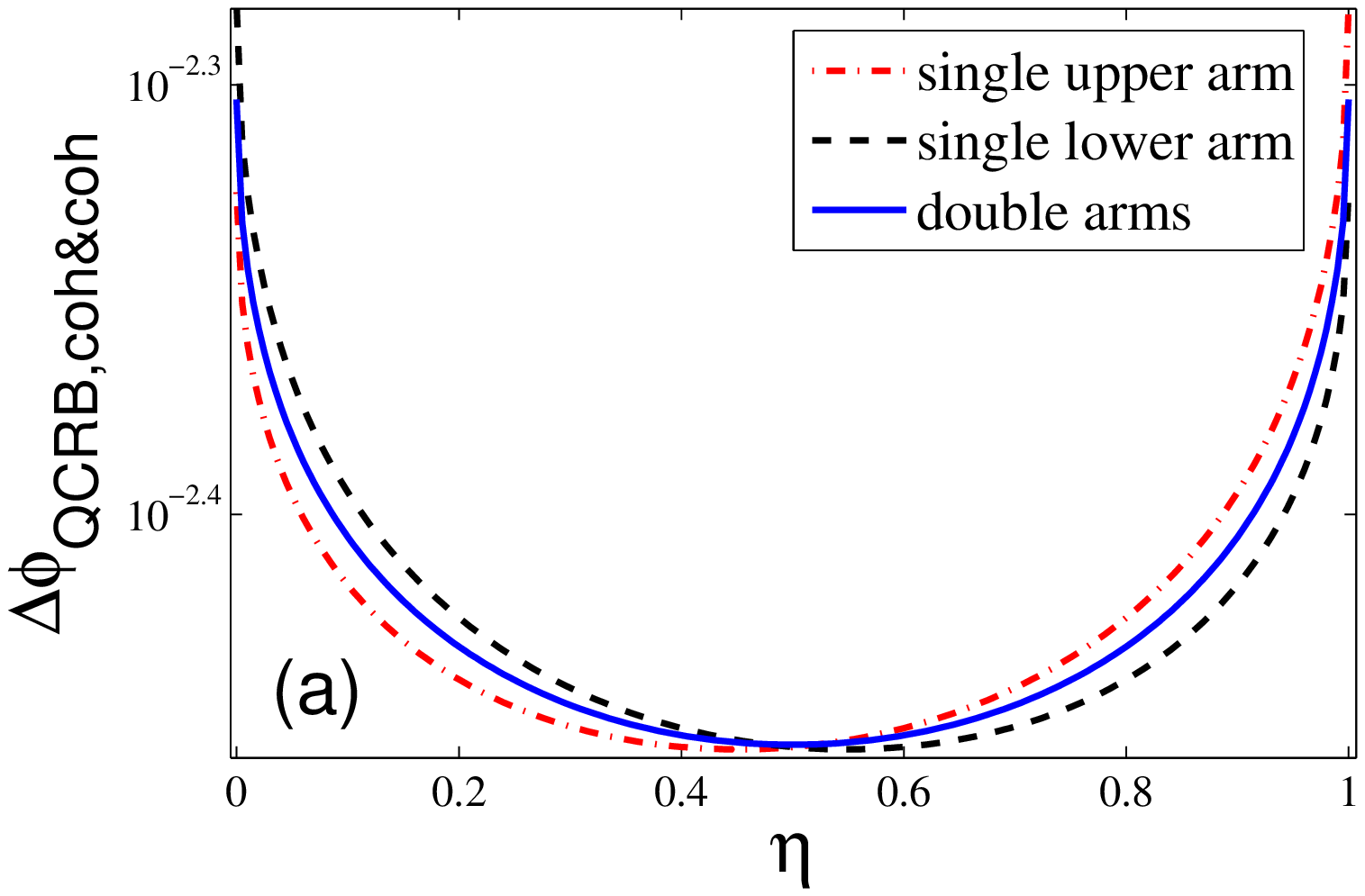}} %
\centerline{\includegraphics[scale=0.5,angle=0]{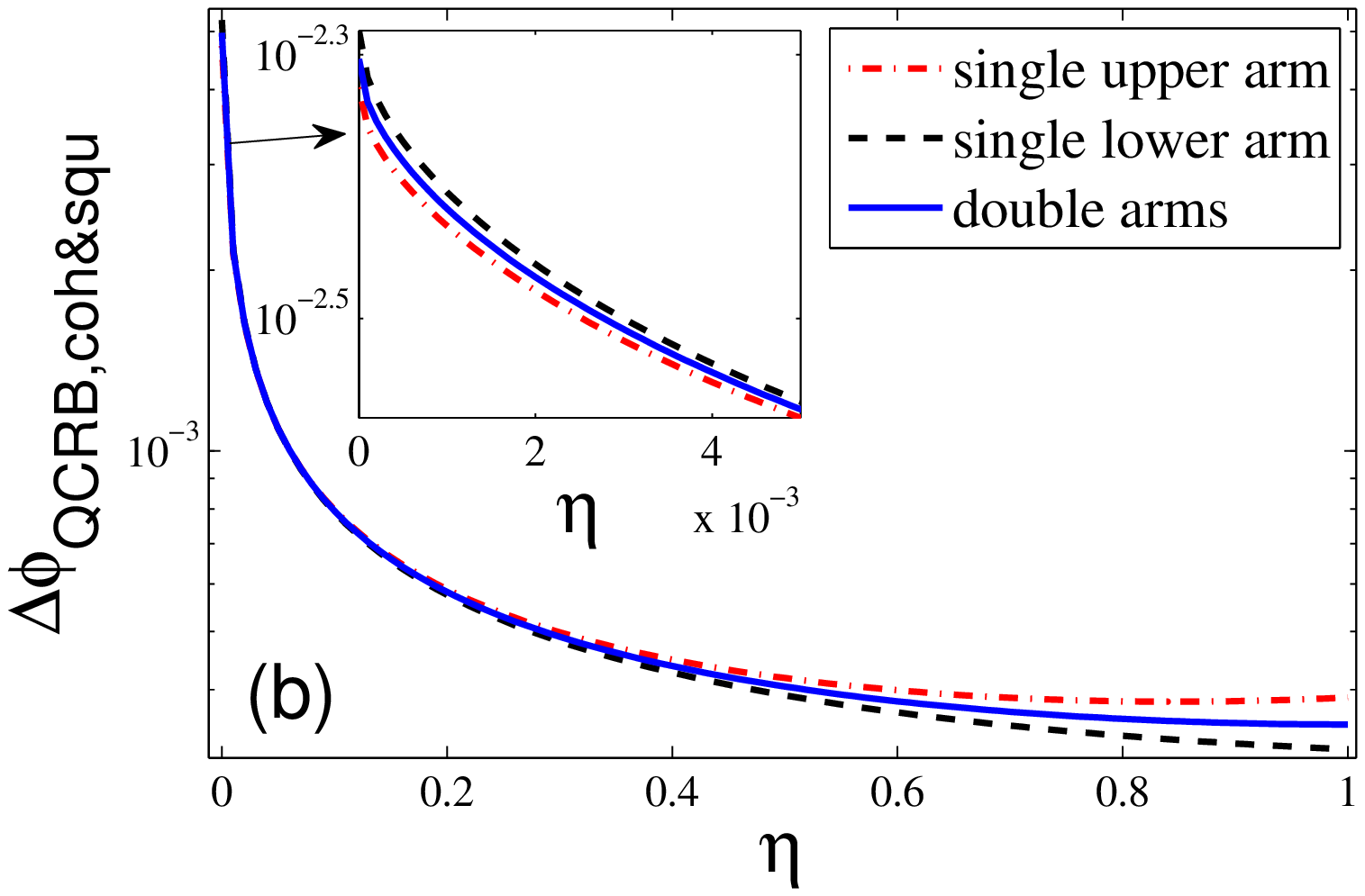}}
\caption{The QCRB versus the $\protect\eta$ for (a) two coherent states
input; (b) coherent $\otimes$ squeezed state input. The inset in figure (b)
shows a zoom of the graph for small values of $\protect\eta$. Parameters: $%
N_{\mathrm{in}}=200$, and $g=1.5$.}
\label{fig3}
\end{figure}

Whatever the measurement chosen, the QCRB can give the lower bound for the
phase measurement \cite{Braunstein94,Braunstein96,Toth, PezzBook, Demkowicz}%
\begin{equation}
\Delta \phi _{\mathrm{QCRB}}=\frac{1}{\sqrt{\mathcal{F}}}.  \label{QCRB}
\end{equation}%
To describe the effect on the QCRB from the unbalanced input state, we
introduce a parameter $\eta $ which is defined by \cite{Hu}
\begin{equation}
\eta =\frac{\mathrm{mean~photon~number~of~}b\mathrm{~mode}}{\mathrm{%
total~mean~photon~number~of~input}}.
\end{equation}%
For the two coherent states input, $\eta $ is equal to $N_{\beta }/N_{%
\mathrm{in}}$ ($N_{\mathrm{in}}=N_{\alpha }+N_{\beta }$), and the optimal
phase sensitivities $\Delta \phi _{\mathrm{QCRB}}$ as a function of $\eta $\
are shown in Fig.~3(a). When $\eta $ is small, the $\Delta \phi _{\mathrm{%
QCRB}}$ from the single upper arm case is the best. But when $\eta $ is
large, the $\Delta \phi _{\mathrm{QCRB}}$ from the single lower\ arm case is
the best, and the $\Delta \phi _{\mathrm{QCRB}}$ from the two-arm case is
always an intermediate value. For a given fixed $N_{\mathrm{in}}$, and the
two coherent states input case, the optimal value $\eta $ is $0.5$. That is
for the two coherent states input the optimal input state is $|\sqrt{N_{%
\mathrm{in}}/2}e^{i\theta _{\alpha }}\rangle \otimes |-\sqrt{N_{\mathrm{in}%
}/2}e^{-i\theta _{\alpha }}e^{i\theta _{g}}\rangle $, and the corresponding
optimal QFI is $\mathcal{F}_{\mathrm{coh\&coh}}^{\mathrm{T,opt}}=N_{\mathrm{%
in}}e^{4g}+\sinh ^{2}(2g)$. The optimal QFI $\mathcal{F}_{\mathrm{coh\&coh}%
}^{\mathrm{T,opt}}$ as a function of the total input mean photon number $N_{%
\mathrm{in}}$\ is shown in Fig. 4 (the blue dot-dashed line).

\begin{figure}[ptbh]
\centerline{\includegraphics[scale=0.5,angle=0]{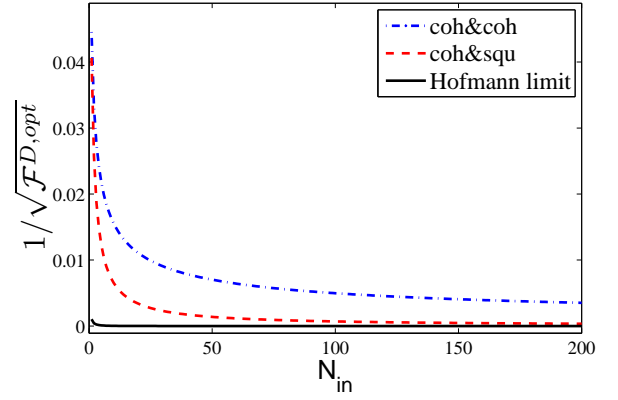}}
\caption{The optimal QFIs versus the total input mean photon number $N_{%
\mathrm{in}}$. The dot-dashed line and the dashed line are the two coherent
states input and coherent $\otimes$ squeezed state input, respectively. The
solid line is the Hofmann limit with coherent $\otimes$ squeezed state
input. $g=1.5$.}
\label{fig4}
\end{figure}

For coherent $\otimes $ squeezed vacuum state input, $\eta $ is equal to $%
\sinh ^{2}r/N_{\mathrm{in}}$ ($N_{\mathrm{in}}=N_{\alpha }+\sinh ^{2}r$),
where the parameter $\eta $ can be used to label the squeezing fraction of
the mean photon number. When $\eta =0$ or $\eta =1$, the input state is only
a coherent state $|\alpha \rangle _{a}$ or only a squeezed vacuum state $%
|0,\zeta \rangle _{b}$. When $0<\eta <1$, the input state is a coherent $%
\otimes $ squeezed vacuum state. For coherent $\otimes $ squeezed vacuum
state input case, only the squeezed vacuum light as input and without the
coherent state, the phase sensitivity is the highest shown in Fig.~3(b).
That is the optimal input state is $|0\rangle \otimes |0,\zeta \rangle $,
and the corresponding optimal QFI is $\mathcal{F}_{\mathrm{coh\&squ}}^{%
\mathrm{T,opt}}=(1+N_{\mathrm{in}})[2N_{\mathrm{in}}\cosh ^{2}(2g)+\sinh
^{2}(2g)]$, which is different from the commonly used optimal input state
with $\left\vert \alpha \right\vert ^{2}\simeq \sinh ^{2}(r)\simeq N_{%
\mathrm{in}}/2$ in MZI \cite{Pezz08,Lang13,Liu13}. The reason is the number
fluctuations and Pasquale \emph{et al.} \cite{Pasquale} have given the same
result for generic two-mode interferometric setup recently. The optimal QFI $%
\mathcal{F}_{\mathrm{coh\&squ}}^{\mathrm{T,opt}}$ as a function of $N_{%
\mathrm{in}}$\ is shown in Fig. 4 (the red dashed line). For a fixed mean
photon number (with number fluctuations), Hofmann suggested the form of
Heisenberg limit is $1/\langle \hat{N}^{2}\rangle ^{1/2}$, which indicates
averaging over the squared photon numbers \cite{hofmann}. In our proposal $%
\langle \hat{N}\rangle $ is defined as $\langle \Psi |(\hat{n}_{a}+\hat{n}%
_{b})|\Psi \rangle $. In Fig. 4 the black solid line is the Hofman limit for
coherent $\otimes $ squeezed vacuum state input under the optimal condition.

For the lossy interferometers, the pure states evolve into the mixed states
and the QFI will be reduced. However, the QFI of pure state puts an upper
bound on that of mixed state. Here, we focus on the maximal QFI of the
SU(1,1) interferometer, then we ignore the losses in the interferometer.

\section{Conclusion}

In conclusion, the analytical expressions of QFI for an SU(1,1)
interferometer with two coherent states and coherent $\otimes $ squeezed
vacuum state inputs have been derived. For single-arm case, the QCRBs of
phase shift in upper arm and in lower arm are slightly different because the
intensities in two interferometric arms are asymmetric. The phase
sensitivities of phase shifts between the single-arm case and two-arm case
are also compared. The QCRB of single-arm case can be slightly higher or
lower than that of two-arm case, which depends on the intensities of the two
arms of the interferometer. For coherent state $\otimes$ squeezed vacuum
state input with a definite input number of photons, the optimal condition
to obtain the highest phase sensitivity is a squeezed vacuum in one mode and
the vacuum state in the other mode.

\begin{widetext}
\begin{table}[h]
\caption{The maximal QFIs of the SU(1,1) interferometer for different phase
delay ways with different input states}
\label{Tab1}\tabcolsep 0.5mm \doublerulesep 2mm
\par
\begin{center}
\renewcommand\arraystretch{1.5} \noindent%
\begin{tabular}{|l|l|l|l|}
\hline
& \multicolumn{2}{|l|}{\ \ \ \ \ \ \ \ \ \ \ \ \ \ \ \ \ \ \ \ \ \ \ \ \ \ \
\ \ \ \ single arm $\mathcal{F}^{\mathrm{S}}$} & \ \ phase shifts in \\
\cline{2-3}
\ \ \ input states & \ \ \ phase shift in upper arm & \ \ \ phase shift in lower arm &
\ \ \  two arms $\mathcal{F}^{\mathrm{T}}$ \\ \hline
two coherent states & $(N_{\alpha }+N_{\beta })\cosh 4g+\sinh ^{2}(2g)$ & $%
(N_{\alpha }+N_{\beta })\cosh 4g+\sinh ^{2}(2g)$ & $(N_{\alpha }+N_{\beta
})\cosh 4g$ \\
$|\alpha \rangle _{a}\otimes |\beta \rangle _{b}$ & $+2\sqrt{N_{\alpha
}N_{\beta }}\sinh 4g+N_{\alpha }+N_{\beta }$ & $+2\sqrt{N_{\alpha }N_{\beta }%
}\sinh 4g+N_{\alpha }+N_{\beta }$ & $\ +\sinh ^{2}(2g)$ \\
& $+2(N_{\alpha }-N_{\beta })\cosh 2g$$^{\mathrm{a}}$ & $-2(N_{\alpha
}-N_{\beta })\cosh 2g$ & $+2\sqrt{N_{\alpha }N_{\beta }}\sinh 4g$ \\ \hline
coherent $\otimes $ squee & $\cosh ^{2}(2g)[\sinh ^{2}(2r)/2+N_{\alpha }]$ &
$\cosh ^{2}(2g)[\sinh ^{2}(2r)/2+N_{\alpha }]$ & $\cosh ^{2}(2g)[N_{\alpha }$
\\
-zed vacuum states & $+\sinh ^{2}(2g)[N_{\alpha }e^{2r}+\cosh ^{2}r]$ & $%
+\sinh ^{2}(2g)[N_{\alpha }e^{2r}+\cosh ^{2}r]$ & $+\sinh ^{2}(2r)/2]$ \\
$|\alpha \rangle _{a}\otimes |\varsigma ,0\rangle _{b}$ & $+N_{\alpha
}(1+2\cosh 2g)$ & $+N_{\alpha }(1-2\cosh 2g)$ & $+\sinh ^{2}(2g)[N_{\alpha
}e^{2r}$ \\
$\left( \varsigma =r\exp (i\theta _{\varsigma })\right) $ & $-\frac{1}{4}%
(\cosh 4r-1)(2\cosh 2g-1)^{\mathrm{b}}$ & $+\frac{1}{4}(\cosh 4r-1)(2\cosh
2g+1)^{\mathrm{b}}$ & $+\cosh ^{2}r]$ \\ \hline
\multicolumn{4}{l}{$^{\mathrm{a}}${Ref.~ \cite{Sparaciari}}.} \\
\multicolumn{4}{l}{$^{\mathrm{b}}${Ref.~ \cite{Li2016}}.}%
\end{tabular}%
\end{center}
\end{table}
\end{widetext}
\section*{Acknowledgements}

This work was supported by the National Natural Science Foundation of China
under Grant Nos.~11474095, 11654005 and 11234003, and the National Key
Research and Development Program of China under Grant No.~2016YFA0302000.

\end{document}